\documentclass[pra,aps,superscriptaddress,twocolumn]{revtex4-1}

\usepackage[colorlinks,bookmarks=false,citecolor=magenta,linkcolor=magenta,urlcolor=
magenta]{hyperref}
\usepackage{amsmath,amssymb}
\usepackage{setspace}
\usepackage{graphicx}
\usepackage{float}
\usepackage{dcolumn}
\usepackage{bm}
\usepackage{subfigure}
\usepackage{pifont} 
\usepackage{braket}
\usepackage{siunitx}
\usepackage{booktabs}
\usepackage{array}
\usepackage{natbib}
\usepackage{multirow}
\usepackage{ulem}
\usepackage{changes}
\definechangesauthor[name={Xinhui Ruan}, color=blue]{rxh}
\definechangesauthor[name={Li Li}, color=blue]{Li}

\begin{document}

\preprint{APS/123-QED}

\title{Quantum Acoustics with Superconducting Qubits in the Multimode Transition-Coupling Regime}

\author{Li Li}
\thanks{These authors contributed equally to this work.}
\affiliation{
Beijing National Laboratory for Condensed Matter Physics,
Institute of Physics, Chinese Academy of Sciences, Beijing 100190, China
}
\affiliation{
School of Physical Sciences, University of Chinese Academy of Sciences, Beijing 100049, China
}

\author{Xinhui Ruan}
\thanks{These authors contributed equally to this work.}
\affiliation{
Beijing National Laboratory for Condensed Matter Physics,
Institute of Physics, Chinese Academy of Sciences, Beijing 100190, China
}
\affiliation{
Department of Automation, Tsinghua University, Beijing 100084, P. R. China
}
\affiliation{
	Key Laboratory of Low-Dimensional Quantum Structures and Quantum Control of Ministry of Education, Key Laboratory for Matter Microstructure and Function of Hunan Province, Department of Physics and Synergetic Innovation Center for Quantum Effects and Applications, Hunan Normal University, Changsha 410081, People’s Republic of China
}

\author{Si-Lu Zhao}
\affiliation{
Beijing National Laboratory for Condensed Matter Physics, 
Institute of Physics, Chinese Academy of Sciences, Beijing 100190, China
}
\affiliation{
School of Physical Sciences, University of Chinese Academy of Sciences, Beijing 100049, China
}

\author{Bing-Jie Chen}
\affiliation{
Beijing National Laboratory for Condensed Matter Physics, 
Institute of Physics, Chinese Academy of Sciences, Beijing 100190, China
}
\affiliation{
School of Physical Sciences, University of Chinese Academy of Sciences, Beijing 100049, China
}

\author{Gui-Han Liang}
\affiliation{
Beijing National Laboratory for Condensed Matter Physics, 
Institute of Physics, Chinese Academy of Sciences, Beijing 100190, China
}
\affiliation{
School of Physical Sciences, University of Chinese Academy of Sciences, Beijing 100049, China
}

\author{Yu Liu}
\affiliation{
Beijing National Laboratory for Condensed Matter Physics, 
Institute of Physics, Chinese Academy of Sciences, Beijing 100190, China
}
\affiliation{
School of Physical Sciences, University of Chinese Academy of Sciences, Beijing 100049, China
}

\author{Cheng-Lin Deng}
\affiliation{
Beijing National Laboratory for Condensed Matter Physics, 
Institute of Physics, Chinese Academy of Sciences, Beijing 100190, China
}
\affiliation{
School of Physical Sciences, University of Chinese Academy of Sciences, Beijing 100049, China
}

\author{Wei-Ping Yuan}
\affiliation{
Beijing National Laboratory for Condensed Matter Physics, 
Institute of Physics, Chinese Academy of Sciences, Beijing 100190, China
}
\affiliation{
School of Physical Sciences, University of Chinese Academy of Sciences, Beijing 100049, China
}
\author{Jia-Cheng Song}
\affiliation{
Beijing National Laboratory for Condensed Matter Physics, 
Institute of Physics, Chinese Academy of Sciences, Beijing 100190, China
}
\affiliation{
School of Physical Sciences, University of Chinese Academy of Sciences, Beijing 100049, China
}

\author{Zheng-He Liu}
\affiliation{
Beijing National Laboratory for Condensed Matter Physics, 
Institute of Physics, Chinese Academy of Sciences, Beijing 100190, China
}
\affiliation{
School of Physical Sciences, University of Chinese Academy of Sciences, Beijing 100049, China
}

\author{Tian-Ming Li}
\affiliation{
Beijing National Laboratory for Condensed Matter Physics, 
Institute of Physics, Chinese Academy of Sciences, Beijing 100190, China
}
\affiliation{
School of Physical Sciences, University of Chinese Academy of Sciences, Beijing 100049, China
}

\author{Yun-Hao Shi}
\affiliation{
Beijing National Laboratory for Condensed Matter Physics, 
Institute of Physics, Chinese Academy of Sciences, Beijing 100190, China
}

\author{He Zhang}
\affiliation{
Beijing National Laboratory for Condensed Matter Physics,
Institute of Physics, Chinese Academy of Sciences, Beijing 100190, China
}
\affiliation{
School of Physical Sciences, University of Chinese Academy of Sciences, Beijing 100049, China
}

\author{Ming Han}
\affiliation{
Beijing National Laboratory for Condensed Matter Physics,
Institute of Physics, Chinese Academy of Sciences, Beijing 100190, China
}
\affiliation{
School of Physical Sciences, University of Chinese Academy of Sciences, Beijing 100049, China
}

\author{Jin-Ming Guo}
\affiliation{
Beijing National Laboratory for Condensed Matter Physics,
Institute of Physics, Chinese Academy of Sciences, Beijing 100190, China
}
\affiliation{
School of Physical Sciences, University of Chinese Academy of Sciences, Beijing 100049, China
}
\author{Xue-Yi Guo}
\affiliation{
Beijing Academy of Quantum Information Sciences, Beijing 100193, China
}

\author{Qianchuan Zhao}
\affiliation{
Department of Automation, Tsinghua University, Beijing 100084, P. R. China
}

\author{Jing Zhang}
\affiliation{School of Automation Science and Engineering, Xi’an Jiaotong University, Xi’an 710049, China}
\affiliation{MOE Key Lab for Intelligent Networks and Network Security, Xi’an Jiaotong University, Xi’an 710049, China}

\author{Pengtao Song}
\affiliation{School of Automation Science and Engineering, Xi’an Jiaotong University, Xi’an 710049, China}
\affiliation{MOE Key Lab for Intelligent Networks and Network Security, Xi’an Jiaotong University, Xi’an 710049, China}

\author{Xiaohui Song}
\affiliation{
Beijing National Laboratory for Condensed Matter Physics, 
Institute of Physics, Chinese Academy of Sciences, Beijing 100190, China
}
\affiliation{
Hefei National Laboratory, Hefei 230088, China
}

\author{Kai Xu}

\affiliation{
Beijing National Laboratory for Condensed Matter Physics, 
Institute of Physics, Chinese Academy of Sciences, Beijing 100190, China
}
\affiliation{
Beijing Academy of Quantum Information Sciences, Beijing 100193, China
}
\affiliation{
Hefei National Laboratory, Hefei 230088, China
}

\author{Heng Fan}

\affiliation{
Beijing National Laboratory for Condensed Matter Physics, 
Institute of Physics, Chinese Academy of Sciences, Beijing 100190, China
}
\affiliation{
School of Physical Sciences, University of Chinese Academy of Sciences, Beijing 100049, China
}
\affiliation{
Beijing Academy of Quantum Information Sciences, Beijing 100193, China
}
\affiliation{
Hefei National Laboratory, Hefei 230088, China
}

\author{Yu-Xi Liu}

\affiliation{School of Integrated Circuits,Tsinghua University, Beijing 100084, China
}
\author{Zhihui Peng}
\email{zhihui.peng@hunnu.edu.cn}
\affiliation{
	Key Laboratory of Low-Dimensional Quantum Structures and Quantum Control of Ministry of Education, Key Laboratory for Matter Microstructure and Function of Hunan Province, Department of Physics and Synergetic Innovation Center for Quantum Effects and Applications, Hunan Normal University, Changsha 410081, People’s Republic of China
}
\affiliation{
Hefei National Laboratory, Hefei 230088, China
}

\author{Zhongcheng Xiang}
\email{zcxiang@iphy.ac.cn}

\affiliation{
Beijing National Laboratory for Condensed Matter Physics, 
Institute of Physics, Chinese Academy of Sciences, Beijing 100190, China
}
\affiliation{
Hefei National Laboratory, Hefei 230088, China
}

\author{Dongning Zheng}
\email{dzheng@iphy.ac.cn}
\affiliation{
Beijing National Laboratory for Condensed Matter Physics, 
Institute of Physics, Chinese Academy of Sciences, Beijing 100190, China
}
\affiliation{
School of Physical Sciences, University of Chinese Academy of Sciences, Beijing 100049, China
}
\affiliation{
Hefei National Laboratory, Hefei 230088, China
}

\date{\today}

\begin{abstract}
Hybrid mechanical-superconducting systems for quantum information processing have attracted significant attention due to their potential applications. In such systems, the weak coupling regime, dominated by dissipation, has been extensively studied. The strong coupling regime, where coherent energy exchange exceeds losses, has also been widely explored. However, the transition-coupling regime, which lies between the above two and exhibits rich, unique physics, remains underexplored. In this study, we fabricate a tunable coupling device to investigate the coupling of a superconducting transmon qubit to a seven-mode surface acoustic wave resonator (SAWR), with a particular focus on the transition-coupling regime. Through a series of phonon oscillation experiments and studies in the dispersive regime, we systematically characterize the performance of the SAWR. We then explore the complex dynamics of energy exchange between the qubit and the mechanical modes, highlighting the interplay between dissipation and coherence. Finally, we propose a protocol for qubit readout and fast reset with a multimode mechanical cavity using one mode for readout and another mode for reset. We have demonstrated in simulation that the qubit achieves both fast reset and high coherence performance when the qubit is coupled to the reset mode in the transition-coupling regime.
\end{abstract}

\maketitle


\section{Introduction\label{sec1}}

Quantum acoustics, the study of quantum phenomena in phononic systems, is rapidly advancing with applications in quantum information processing~\cite{gustafsson2012,chu2017,qiao2023}. Phononic systems offer advantages including longer coherence times~\cite{manenti2016,maccabe2020a,iyer2024,liu2025} and higher mode density~\cite{moores2018}. These properties enable diverse applications, such as microwave-optical conversion~\cite{bochmann2013,andrews2014,bagci2014,shumeiko2016,mirhosseini2020}, quantum registers~\cite{palomaki2013}, microwave amplifiers~\cite{massel2011}, and feedback networks~\cite{kerckhoff2013}, highlighting the potential of phononic systems in next-generation quantum technologies. Recently, various hybrid quantum acoustic systems have been explored, including mechanical oscillator ~\cite{teufel2011,pirkkalainen2013,rouxinol2016}, surface acoustic wave resonator ~\cite{manenti2017,noguchi2017,bolgar2018}, and bulk acoustic wave resonator ~\cite{chu2018,vonlupke2022,lupke2024}. These systems have been coupled to superconducting circuits~\cite{bild2023}, optomechanical setups~\cite{kline2024}, and spin systems~\cite{thomas2021}, providing versatile platforms for studying quantum behaviors through phononic interactions. Among them, circuit quantum acoustodynamics~(cQAD), which involves coupling superconducting qubits with acoustic resonators, is an important platform in the field. This approach combines the coherence of phononic systems with the high coherence~\cite{wang2022a}, controllability~\cite{heunisch2023a}, and scalability~\cite{liang2023} of superconducting qubits, making cQAD systems promising candidates for quantum state transfer~\cite{dumur2021,bienfait2019} and quantum random access memories~\cite{hann2019,wang2024d}.

In cQAD, the coupling between superconducting qubits and mechanical resonators has allowed researchers to explore the quantum mechanical properties of the phononic system. Earlier studies concentrated predominantly on the weak-coupling or strong-coupling regimes, where the dynamics is well-established and well-understood. The weak-coupling regime enables the study of phonon-mediated decoherence~\cite{ioffe2004,kline2024,cleland2024}. In contrast, the strong-coupling regime enables coherent energy exchange between qubits and mechanical resonators, thus allowing the control of quantum acoustic states~\cite{chu2018,bild2023} and qubit-driven phonon interactions~\cite{lupke2024,wei2024}. However, the intermediate regime, also known as transition-coupling regime in cQAD, which bridges the weak- and strong-coupling limits, has remained largely unexplored. This regime is particularly intriguing, as it enables the exploration of rich and complex dynamics, where both dissipative effects and coherent interactions come into play. For example, a quantum emitter in the transition-coupling regime will exhibit an additional bandwidth modulation degree of freedom~\cite{liberal2019}. On the other hand, by utilizing the properties of transition-coupling regime to balance coherent coupling and dissipation, the sensitivity of quantum sensing can be enhanced~\cite{sensor1,sensor2}. Furthermore, the physics of exceptional points (EPs) based on transition-coupling regime continues to be applied in various domains of non-Hermitian physics~\cite{zhong2020,choi2017,hodaei2014,song2024,zhang2022}. Researches on the transition-coupling regime and its implementation in cQAD are highly meaningful.
\begin{figure*}[t]
    \centering
    \includegraphics[height = 9.5 cm]{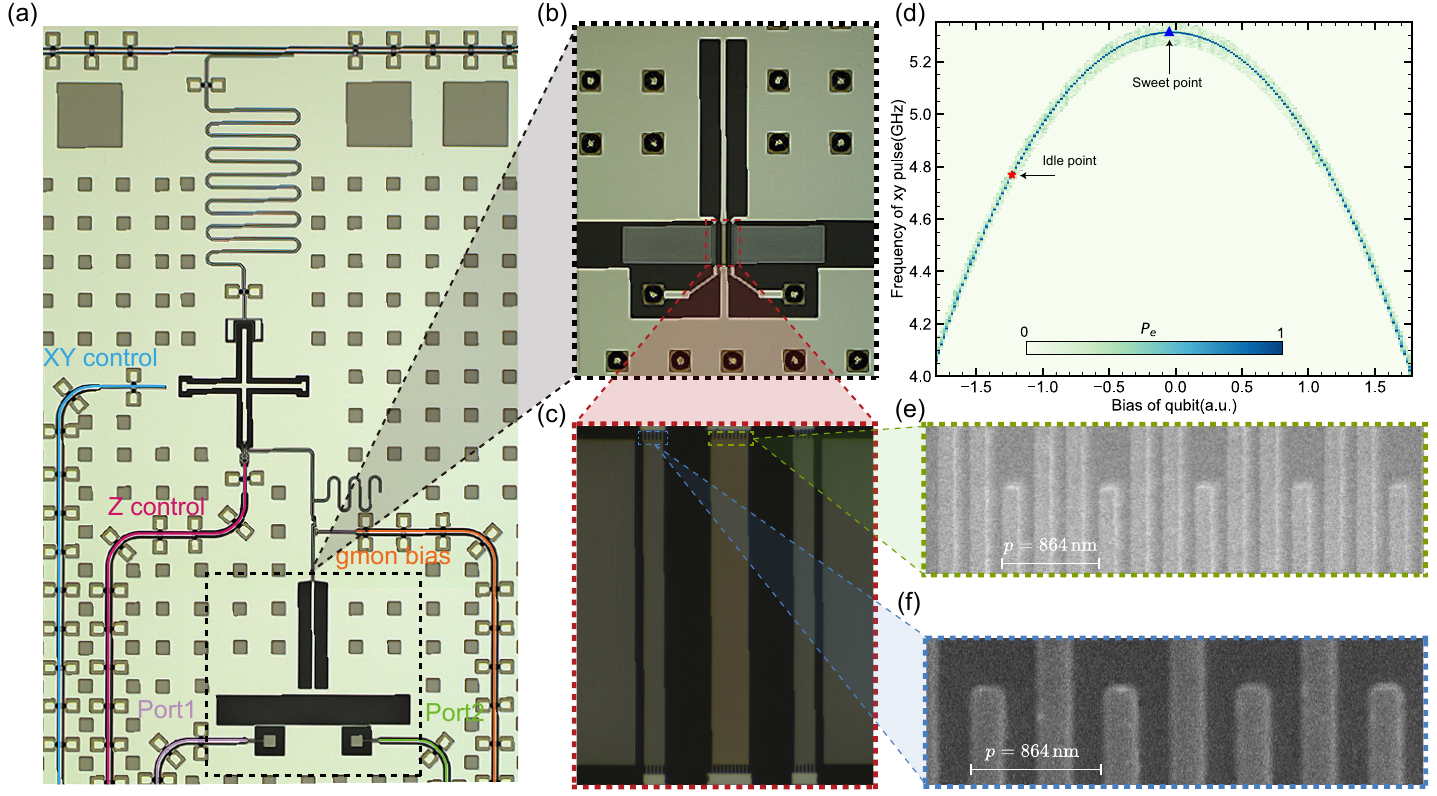}
    \caption{The device. (a) Optical microscope image of the bottom sapphire chip, where the superconducting quantum circuits are fabricated. The black dashed box highlights the region containing the SAWR. (b) Enlarged view of the black dashed box in panel (a), showing the optical microscope image of the SAWR, which is fabricated on the top $\text{LiNbO}_3$ chip. The SAWR includes three interdigital transducer (IDT) ports. (c) Magnified view of the red box in panel (b), providing an optical microscope image of the three IDTs. (d) Measured transition frequency of the transmon qubit as a function of bias voltage. (e) Enlarged view of the green box in panel (c), showing the SEM image of the middle IDT, which has a period $p$ and a finger width of $d_I=p/6=$\SI{144}{nm}. (f) Enlarged view of the blue box in panel (c), showing a scanning electron microscope (SEM) image of the left and right IDTs, which serve as input and output transducers. These IDTs are connected to ports 1 and 2 on the bottom chip via indium bumps. Each finger has a width of $d_s=$\SI{216}{nm}, and the IDT period is $p=4d_s=$\SI{864}{nm}.} 
    \label{fig1}
\end{figure*}

In this study, we employed flip-chip technologies to fabricate a superconducting transmon qubit and a seven-mode SAWR with a tunable coupling between them. We began with detailed microstructural measurements by phonons oscillation experiment to verify the integrity and alignment of the bonded device, ensuring acoustic coupling between the qubit and the SAWR. Subsequently, we examined the AC Stark effect within the dispersive regime to characterize phononic interactions and probed the decay rate of individual mechanical modes. Finally, we extended our investigation to the transition-coupling regime, where resonant evolution allowed us to explore the dynamics of energy exchange between the qubit and phononic modes. We observe a transition from underdamped behavior with coherent oscillations to overdamped behavior characterized by decay without oscillations, depending on the mode number and the coupler bias. We demonstrated that a multimode cavity enables both dispersive readout and fast reset of a qubit. When the fast reset mode is in the transition-coupling regime with the qubit, high reset efficiency is achieved while the induced qubit decoherence remains negligible. This approach not only highlights the practical utility of the transition-coupling regime but also provides new insights for optical readout, contributing to the design of future large-scale superconducting quantum chips. By systematically studying the dynamics that depend on both the modes and the coupler, our work provides insights into the interplay between dissipative and coherent interactions in quantum acoustics, thereby advancing the development of robust and tunable quantum systems.

\section{The device\label{sec2}}
In this experiment, we designed a multi-mode SAWR and coupled it to a transmon qubit through flip-chip technology. The transmon qubit and its control lines are fabricated on the sapphire substrate, shown in Fig.~\ref{fig1}(a). The SAWR is a 2D Fabry-Perot cavity formed by IDTs and Bragg mirrors and fabricated on the $128^{\circ}$Y-X $\text{LiNbO}_3$ substrate, as shown in Figs.~\ref{fig1}(b)-(c) and (e)-(f). In order to directly control the surface acoustic wave (SAW) modes, we additionally integrated two input/output IDTs (IDT1 and IDT3 in Fig.~\ref{fig2}(a)) into the SAWR. The middle IDT (IDT2 in Fig.~\ref{fig2}(a)) is used to perform energy conversion between the SAW and the qubit. After alignment and flip-chip process, there is an inter-chip mutual inductance caused by the overlapping between the inductive line of the IDT2 and the corresponding part of the bottom chip. We use a gmon as the coupler to tune the coupling strength between SAWR and the qubit by applied voltage bias through gmon bias line~\cite{chen2014}. The fabrication processes of this device are the same as our former work~\cite{ruan2024}.

The period $p$ of the IDTs is \SI{864}{nm}, corresponding to the centre frequency of the SAWR around \SI{4.6}{GHz}. The length and thick of all the electrodes are \SI{75}{\micro m} and \SI{30}{nm}, respectively. The width of IDT1, 3 and Bragg mirrors is $p/4$, while that of IDT2 is $p/6$ and each cell of IDT2 has three fingers (Fig.~\ref{fig1}(f)). This design aims to minimize the reflection effect of IDT2 on the acoustic wave. The cell numbers of the three IDTs are 5, 10 and 5, respectively, while each mirror has 400 periodic stripe electrodes. The distance between the mirrors is $50\times p= $\SI{50}{\micro m}. Unlike single-IDT resonator, the modes in multi-IDT resonator are more complex. 

The transmon qubit has a DC SQUID and its transition frequency can be tuned through changing the external flux applied to the SQUID. The transition frequency of the qubit under different Z bias is shown in Fig.~\ref{fig1}(d). 

\section{Phononic response of SAWR\label{sec3}}
\begin{figure}[!h]
    \centering
    \includegraphics[height = 12 cm]{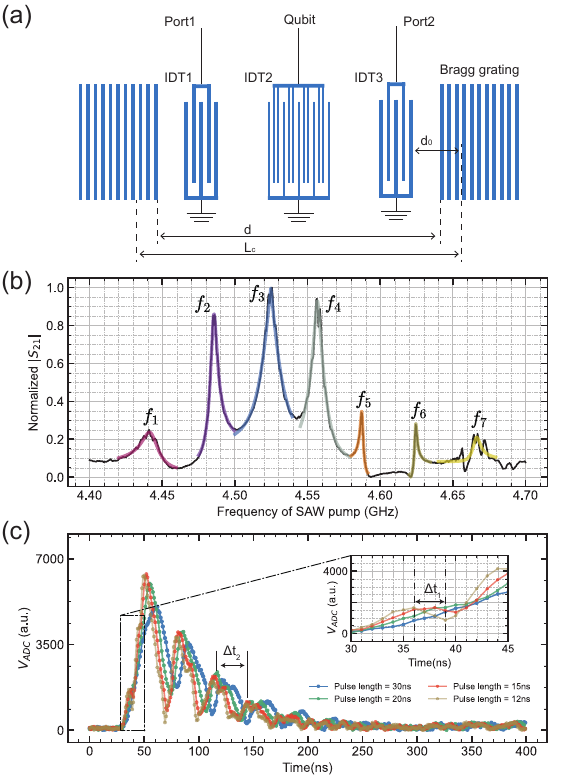}
    \caption{Phononic response of SAWR. (a) Schematic of the SAWR, with the main structure consisting of three IDTs and two Bragg gratings. (b) $S_{21}$ of the SAWR with fits applied to the data from the near-resonant regions of seven peaks. (c) Time-resolved measurement of the SAWR. It is performed by applying a \SI{12}{ns} electrical Gaussian pulse to the Port1 and acquiring the output signal from the Port2. The electrical signals are collected by the Analog-to-Digital Converter (ADC) with a time resolution of $1\,\text{ns}$.}
    \label{fig2}
\end{figure}
  
The schematic of the SAWR is shown in Fig.~\ref{fig2}(a). We conducted measurements in a commercial dilution refrigerator, with a base temperature of around 12\,$\text{mK}$. By performing transmission spectrum measurements across the two IDT ports of the SAWR, we can obtain its resonant frequencies, as illustrated in Fig.~\ref{fig2}(b). The $S_{21}$ spectrum reveals seven distinct peaks. By fitting each peak, the corresponding resonant frequencies, $f_m$, are determined as summarized in Tab.~\ref{tab1}. The non-ideal IDTs caused asymmetric reflection of SAW~\cite{uno1982} leading to irregularities in certain peak profiles. Additionally, impedance mismatch and transmission line response contribute to the asymmetry observed in some peak shapes.

Assuming the device is symmetric, we can calculate the SAWs propagation speed on the $128^{\circ}$Y-X $\text{LiNbO}_3$ substrate as $v_e = pf_{\text{m4}} \approx 3938\,\text{m/s}$, where $f_{\text{m4}} = 4.5576\,\text{GHz}$  is the frequency of the central mechanical mode. To obtain the structural parameters of the SAWR, we conducted a phonon oscillation experiment~\cite{manenti2017} using the central mechanical mode. A short pulse was applied at Port1 $(f = f_{\text{m4}})$, and the time-dependent response of the electrical signal was monitored at Port2. The pulse duration satisfies $t \leq 2L_{\text{c}}/v_e$, where $L_{\text{c}}$ is the effective cavity length of the SAWR. The experimental results are shown in Fig.~\ref{fig2}(c). Due to the propagation and the reflection by the Bragg grating, the SAW signal captured by the IDT at Port2 exhibits oscillatory decay.

To improve the efficiency of microwave-acoustic conversion, Gaussian pulses were employed in the experiment~(Appendix~\ref{Phonon Oscillation}). Pulses of varying lengths were used to obtain the most accurate structural parameters, as indicated by the color transition from blue to brown in Fig.~\ref{fig2}(c). When the pulse length was reduced from $30\,\text{ns}$ to $12\,\text{ns}$, a distinct downward oscillation emerged at the first rising edge of the curve. This oscillation was caused by SAW propagating through IDT2, reflecting off the Bragg grating on the right, and passing through IDT2 again, as illustrated in the inset of Fig.~\ref{fig2}(c).

The ADC sampling rate employed in our experiments was 1 GSa/s, achieving a time resolution of \SI{1}{\nano\second}. The time between the first rising peak and the subsequent falling peak named $\Delta t_1$ was measured to be $3\sim 4\,\text{ns}$, allowing the calculation of $d_0 = d_1 + L_p =  \Delta t_1 v_e/2 = 5.91 \sim 7.88\,$\SI{}{\micro m} where $L_p$ represents the penetration depth and $d_1$ is the distance between $\text{IDT2}$ and the Bragg grating, measured to be \SI{2.2}{\micro m}. Consequently, $L_p$ is determined to be within the range of 3.71$\sim$\SI{5.68}{\micro m}, corresponding to a single-electrode reflectivity of $r_s = p/\left( 4L_p\right)$, which falls between 0.038 and 0.058. These values align well with the design parameters of $L_c =$\SI{4.65}{\micro m} and $r_s = 0.0473$. Furthermore, the spacing between oscillation peaks corresponds to twice of the effective cavity length, such that $L_c = \Delta t_2 v_e/2 = 53.2\,$\SI{}{\micro m} matches the design value of \SI{53.3}{\micro m}. Experimental data for various modes, pulse shapes, and power levels in phonon oscillation experiments are provided in the Appendix~\ref{Phonon Oscillation}.

\begin{table*}[t]
    \centering
    \caption{The parameters of mechanical modes.}
    \renewcommand{\arraystretch}{1.2} 
    \begin{tabular}{ccccccccccc}
        \hline
         & Quantities  & Symbols  & Units & & &  & Values \\
        \hline
        & Mode & m & & 1 &  2  & 3 & 4 & 5 & 6 & 7 \\
        & Resonant Frequency&$f_m $     &  GHz & 4.4413 & 4.4855 & 4.5247 & 4.5576 & 4.5874 & 4.6243 &  4.6660  \\
        & Phonon lifetime&$T_1^m$  &    ns    & 73.5   & 98.8   & 58.3   & 127.6  & 233.1  & 204.2  &  105.8   \\
        & Decay rate&$\kappa_m/2\pi$ & MHz & 2.17   & 1.61   & 2.73   &  1.25  & 0.68   & 0.78   &  1.50    \\
        & Quality factor & $Q_m$   &     & 2050   & 2783   & 1657   &  3654  & 6716   & 5933   &  3102    \\
        \hline 
    \end{tabular}
    \label{tab1}
\end{table*}

\section{Dispersive Acoustic Coupling: The Acoustic AC-stark Effect\label{sec4}}
\begin{figure*}[t]
    \centering
    \includegraphics[scale = 0.85]{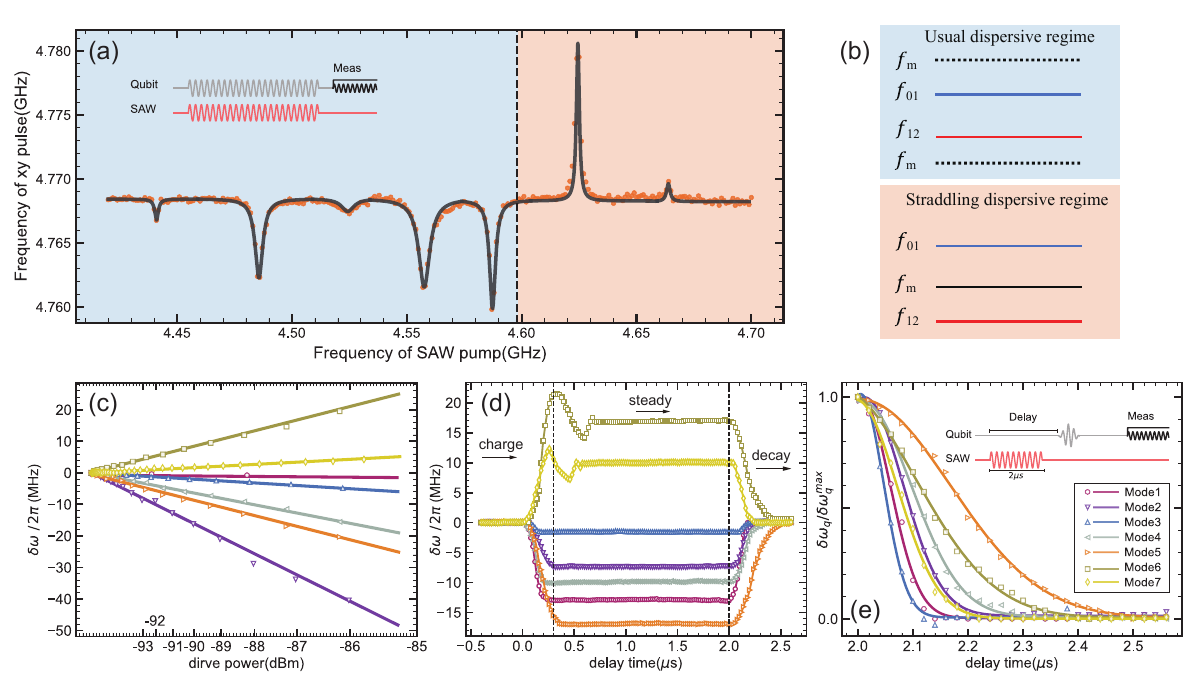}
    \caption{Acoustic AC stark shift measurements in the dispersive regime. The transmon is biased at the idle point and the gmon is set to $V_g = -3.0$. (a) The qubit transition frequency shifts as a function of the SAW pump frequency. The dashed line indicates the transition frequency from the first excited state to the second excited state of the qubit at the idle point, $f_{12} = 4.5979\,\text{GHz}$. (b) Energy level diagrams illustrating the conventional dispersive regime $(\chi < 0)$ and the straddling dispersive regime $(\chi > 0)$. (c) Linear relationship between the qubit transition frequency shift and the SAW pump power. (d) AC Stark shift delay measurements for different SAW pump frequencies. (e) The inset shows the experimental pulse sequence. The main plot presents the normalized data from (d) after $2\,\text{$\mu$s}$, expressed as \(\delta \omega_{\text{q}}/\delta \omega_{\text{q}}^{\text{max}}\). The solid line represents a fit to the data. Colors and line styles in (c), (d), and (e) correspond to different modes.}
    \label{fig3}
\end{figure*}  
In this section, we measure the AC Stark effect of the qubit in the dispersive regime. Initially, the qubit is biased to its idle point via the Z control line, with a transition frequency \( f_{01}^{\text{idle}} = 4.7685 \,\text{GHz} \). Subsequently, the frequencies of the applied pulses for both the qubit and the SAW are scanned simultaneously. When the pump frequency sweeps across the resonant frequencies of the mechanical modes, the qubit, which is dispersively coupled to the resonator, induces an AC Stark shift. This results in a series of Lorentzian profiles in the two-dimensional scan, as depicted in Fig.~\ref{fig3}(a). In the case of multi-mode cavity, the AC Stark shift of the qubit includes contributions from individual modes and cross terms arising from qubit-induced virtual phonon exchanges between non-degenerate modes~\cite{sun2006}. However, under conditions of low phonon numbers and large detuning, the effects of these cross interactions can be neglected~\cite{manenti2017,moores2018,chakram2022}. At the same time, due to the monochromaticity of the pulse frequency, the pump excites phonons only in a single mode during the scan. Drawing an analogy from cQED~\cite{cqed}, the AC Stark shift induced by $n$-phonons in m-th mode is given by $\delta \omega_q \equiv 2\chi_{\text{m}} \langle n\rangle$ where
\begin{align}     
\chi_{\text{m}}= -\frac{g^2E_c/\hbar}{\Delta_{\text{m}} (\Delta_{\text{m}} - E_c/\hbar)},
\end{align}
\( \Delta_{\text{m}}/2\pi = f_{01} - f_m \) , \( E_c/2\pi\hbar \equiv f_{01} - f_{12} = 171\,\text{MHz} \) and $\langle n\rangle$ is the average phonon number in m-th SAW mode. When \( \Delta_{\text{m}} > 0 \) and \( \Delta_{\text{m}} - E_c < 0 \), specifically when \( f_{12} < f_m < f_{01} \), the system is operated in the straddling dispersive regime, resulting in a positive frequency shift~\cite{cqed,koch2007,krantz2019}, as illustrated in Fig.~\ref{fig3}(b). 

By setting the SAW driving frequency to each of the seven resonant frequencies in turn, we varied the driving power and observed a linear relationship between the AC Stark shift and the driving power, as shown in Fig.~\ref{fig3}(c). This relationship enables us to determine the range of driving powers that satisfy the low phonon number condition~\cite{schuster2007a} required for the dispersive regime.

Using the qubit as a probe, we can monitor variations in the phonon number within the resonator by adjusting the delay between the XY pulse and the SAW drive. These variations are reflected in changes to the qubit transition frequency during the SAW drive, as illustrated in Fig.~\ref{fig3}(d). The variation in phonon number occurs in three distinct stages: during the charging phase, the coherent drive rapidly increases the phonon number, which then stabilizes at a steady value due to the balance between pumping and dissipation. Once the drive is removed, the number of phonons undergoes rapid decay. The duration of the drive pulse is $ t_{\text{d}} = 2\,\text{$\mu$s}$ corresponding to the start time of decay. Data collected beyond $t_\text{d}$ are normalized as \(\delta f_{\text{q}}/\delta f_{\text{q}}^{\text{max}}\) for ease of comparison, giving Fig.~\ref{fig3}(e). By fitting these decay curves, we extract the phonon lifetimes, which are summarized in Tab.~\ref{tab1}.

\section{Reasonte Acoustic Coupling: Dynamics in Transition-Coupling Regime}
\label{sec5}
\begin{figure*}[t]
    \centering
    \includegraphics[scale = 0.93]{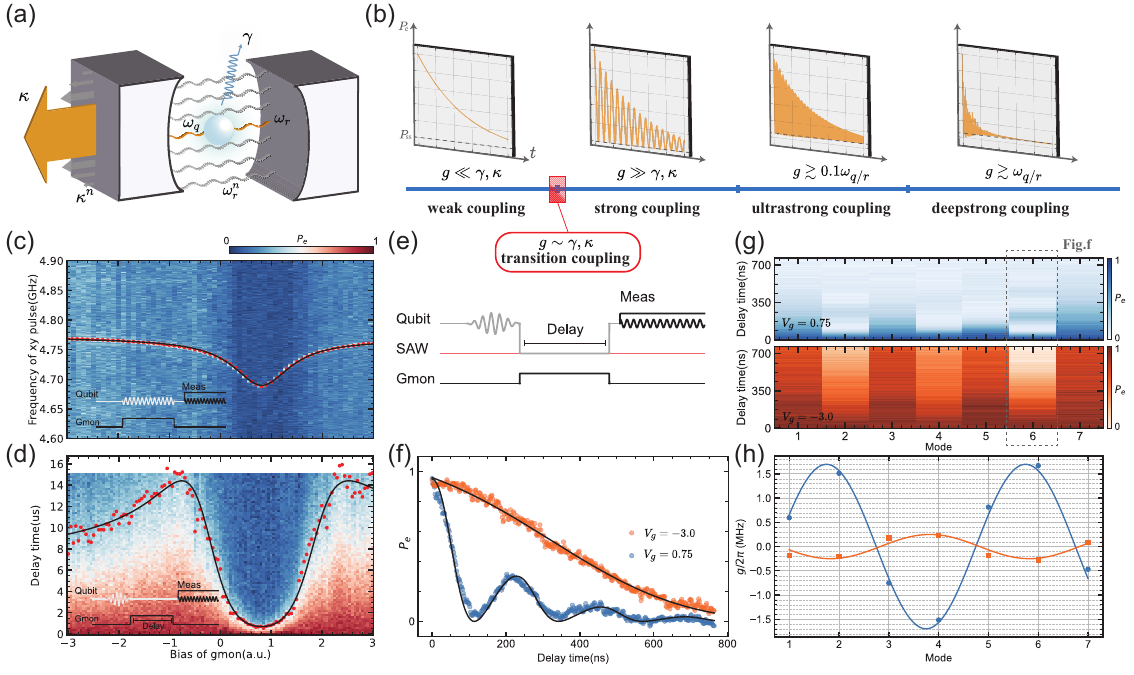}
    \caption{Tunable coupling and resonant regime measurements. (a) Schematic of the model, illustrating a two-level atom coupled to a multimode cavity. The transition frequency of the two-level atom is \( \omega_q = 2\pi f_q \), with an intrinsic dissipation rate \( \gamma \). The frequency of resonant mechanical mode is \( \omega_r = 2\pi f_r  \), with a dissipation rate \( \kappa \). (b) Population dynamics of the $|e\rangle$ state under resonant evolution for different coupling regimes, considering the coupling strength $g$ relative to the dissipation rate and transition frequency. As $g$ increases, the system transitions from the weak coupling regime to the strong, ultrastrong, and deep strong coupling regimes, with each regime exhibiting distinct evolution characteristics. $P_{SS}$ denotes the steady-state population. (c) Dependence of the qubit transition frequency on the gmon bias. Red points indicate frequency values obtained from spectral fitting, while the black solid line represents a fit to these values. (d) Qubit relaxation time $T_1$ of the qubit at the idle point as a function of the gmon bias. Red points denote $T_1$ values extracted from decay measurements, and the black solid line represents a fit to the data. (e) Pulse sequence diagram for the resonant evolution experiment. (f) Resonant evolution data and fits for the qubit interacting with 6th mode, with orange points corresponding to \(V_g = -3.0\) and blue points to \(V_g = 0.75\). (g) Resonant evolution data for seven modes, with orange heatmap for \(V_g = -3.0\) and blue heatmap for \(V_g = 0.75\). (h) Coupling strengths extracted from the fit in (g), exhibiting a sinusoidal dependence on the mode index.}
    \label{fig4}
\end{figure*}

In this section, we examine the qubit dynamics when it is resonant with one of the modes in the multimode cavity. The schematic of the model is shown in Fig.~\ref{fig4}(a). The qubit transition frequency is given by \( \omega_q = 2\pi f_q \), with an intrinsic dissipation rate \( \gamma \), while the resonant mode frequency is \( \omega_r = 2\pi f_r \), with a dissipation rate \( \kappa \). As schematically illustrated in Fig.~\ref{fig4}(b), in the weak coupling regime (\( g \ll \gamma, \kappa \)), the population of the qubit’s excited state undergoes a non-oscillatory decay over time. In the strong coupling regime (\( g \gg \gamma, \kappa \)), pronounced Rabi oscillations emerge, indicating periodic energy exchange between the qubit and the mechanical mode. As the system enters the ultrastrong coupling regime (\( g \gtrsim 0.1\omega_{q/r} \)), the breakdown of the rotating-wave approximation leads to an increased oscillation frequency and an asymmetric waveform. In the deep strong coupling regime (\( g \gtrsim \omega_{q/r} \)), the oscillations become irregular, the system rapidly reaches a steady state, and the steady-state population of the qubit remains nonzero~\cite{gu2017}.

Fig.~\ref{fig4}(c) shows the variation of the qubit transition frequency as a function of the gmon bias. By adjusting the bias of the gmon, both the coupling strength between the qubit and the mechanical modes and the qubit transition frequency are modulated~\cite{geller2014}. We use an in-situ calibration to prevent the qubit frequency from being affected by the gmon bias as described in Appendix~\ref{Frequency Calibration of Qubit}. All AC Stark experiments shown in Fig.~\ref{fig3} were performed with the gmon bias set to $V_g = -3.0$. Fig.~\ref{fig4}(d) shows how the energy relaxation time $T_1$ of the qubit at the idle point varies as a function of the gmon bias.

We model the dissipation of the qubit at the idle point as being influenced by the Purcell effect from seven mechanical modes. The qubit's decay rate is given by~\cite{cqed} 
\begin{equation}
    \gamma_q^{\text{idle}} = \sum_{m=1}^7\left(\frac{g_m}{\Delta_m}\right)^2\kappa_m + \gamma_0^{\text{idle}} ,
    \label{purcell effect}
\end{equation}
where $g_m$ is the coupling strength between the qubit and the mechanical mode-$m$, $\Delta_m/2\pi = f_m - f_q$ and $\gamma_0^{\text{idle}}$ is the intrinsic decay rate of the qubit at the idle point. Fitting the qubit decay data results in the solid black line shown in Fig.~\ref{fig4}(d). As demonstrated in Eq.~\eqref{purcell effect}, the qubit’s decoherence rate is maximized (minimized) when the effective coupling strength between the qubit and the mechanical mode is at its maximum (minimum). This approach allows us to identify both the off point ($V_g = -0.9$) and the maximum coupling point ($V_g = 0.75$) of the gmon coupler, even within the dispersive regime. After decoupling from the mechanical modes, the qubit’s $T_1$ at the idle point increases to approximately $15\,\mu s$. Additionaly, the $T_1$ of the qubit improves significantly across the entire frequency range after decoupling, as shown in Appendix~\ref{Decoherence of qubit}.

Finally, we conducted resonant evolution measurements with $V_g = -3.0$, $V_g = -0.9$ and $V_g = 0.75$, respectively. A $\pi$-pulse was applied to excite the qubit to the $\left| e \right\rangle$ state, followed by a fast Z-pulse to align the qubit's frequency with a specific mode for the evolution during a delay time. The gmon bias was maintained at $V_g$ through its fast Z-control, as illustrated in Fig.~\ref{fig4}(e). The obtained population dynamics of the qubit during resonant evolution with mode-6 are shown in Fig.~\ref{fig4}(f). Figure~\ref{fig4}(g) presents the resonant evolution of the qubit with seven different modes, where the blue heatmap corresponds to \( V_g = 0.75 \) and the orange heatmap corresponds to \( V_g = -3.0 \).

To analyze the data in Figs.~\ref{fig4}(f)-(g), it is essential to accurately understand how the dissipation of the qubit and multiple cavity modes influences the dynamics of the resonant evolution. During this evolution, the behavior of the entire system can be described by the Lindblad master equation
\begin{align}  
    \frac{d\rho}{dt} = & -\frac{i}{\hbar}[ H,\rho ] + \sum_{k = q,m}(2L_k \rho L_k^{\dag} - \{L_k^{\dag}L_k,\rho\}),\\
    \notag
    \frac{H}{\hbar} = & 2\pi f_q b^{\dag}b - \frac{E_c}{2}b^{\dag}b^{\dag}bb \\ 
        & +\sum_{m}[2\pi f_m a_m^{\dag}a_m + g_m(b a_m^{\dag} + b^{\dag} a_m)],\label{master equation hamitonia} \\
    L_q = & \sqrt{\gamma_q}b, L_m = \sqrt{\kappa_m}a_m,
    \label{master equation total}
\end{align}
where $\gamma_q$ is a variable related to the qubit transition frequency. In our experiment, whenever the qubit is resonant with one of the modes, it remains in a dispersive regime with the other six modes. Therefore, the influence of these six detuned modes can be treated as an additive Purcell effect, modeled using a standard dispersive transformation~\cite{temporary-citekey-3099}. Using this insight and truncating Eq.~\eqref{master equation hamitonia} to the two lowest levels of the transmon, the master equation can be equivalently rewritten as:
\begin{align} 
    \notag
    \frac{d\rho}{dt}  = & -\frac{i}{\hbar}[ H_{\text{disp}},\rho ] + (2L_q \rho L_q^{\dag} - \{L_q^{\dag}L_q,\rho\})\\ 
                      & + (2L_m \rho L_q^{\dag} - \{L_m^{\dag}L_m,\rho\}), \\ 
    \notag
    \frac{H_{\text{disp}}}{\hbar}  = & -\frac{2\pi f_q}{2}\sigma_z + 2\pi f_m a_m^{\dag}a_m + g_m(\sigma_{+} a_m^{\dag} + \sigma_{-} a_m)\\
    & + \sum_{n \neq m }\left(2\pi f_n a_n^{\dag}a_n + \chi_n \sigma_z a_n^{\dag}a_n\right),\label{hamitonia of disp} \\ 
    L_q  = &\sqrt{\gamma_q^{\prime}}\sigma_{-}, L_m  = \sqrt{\kappa_m}a_m, 
    \label{master equation disper}
\end{align}
where $\chi_n$ is the dispersive shift of the qubit caused by the $n$-th mode and $\gamma_q^{\prime}$ includes the Purcell effect from the remaining six modes:
\begin{equation}
    \gamma_q^{\prime} = \gamma_q + \sum_{n\neq m}\left(\frac{g_n}{\Delta_n}\right)^2\kappa_n.
\label{qubit gamma}
\end{equation}

Through the steps outlined above, we have incorporated the influence of the non-resonant modes. Next, we will consider the dynamics of resonant evolution by considering the frequency resonance conditions $f_q = f_m$. This is done by performing a rotating frame transformation on Eq.~\eqref{hamitonia of disp} and applying the rotating wave approximation, yielding:
\begin{align}
    \frac{H_{\text{disp}}^{\text{RWA}}}{\hbar}  = g_m(\sigma_{+} a_m^{\dag} + \sigma_{-} a_m),
\end{align}
which gives the standard Jaynes-Cummings interaction Hamiltonian. Focusing only on the two lowest energy levels of the mode, the Hamiltonian in the single-excitation subspace is
\begin{align}
    \frac{H^{\prime}}{\hbar} = g_m(\ket{e,0}_m\bra{g,1}_m + \ket{g,1}_m\bra{e,0}_m),
\end{align}
where $\ket{e}(\ket{g})$ denotes the first excited (ground) state of the qubit, and $\ket{n}_m$ denotes the $n$ photon Fock state of the mode-$m$. By substituting this Hamiltonian into the master equation in the interaction picture, we can derive the expression for the density matrix $\rho_S$. The probability of the qubit being in the $\ket{e}$ state can be written as:
\begin{align}
\notag
P_e(t) & =  \text{Tr}[\rho_S(t)\ket{e}\bra{e}] = f(t)\cdot\text{cos}(g_mt)^2, \\
f(t)   & = \text{exp}\left[ -t\left(\frac{\gamma_q^{\prime} + \kappa_m}{2}\right) + \left(\frac{\kappa_m - \gamma_q^{\prime}}{4g_m}\right)\text{sin}(2g_mt)\right].
\label{P1 of qubit}
\end{align}
\begin{table*}[t]
    \centering
    \caption{The parameters of the qubit under different gmon biases. The $\gamma_q~(\gamma_q^{\prime})$ represents the intrinsic (total) decay rate of the qubit at the transition frequency $f_m$. The $g_m$ represents the coupling strength between the qubit and the $m$-th mechanical mode.}
    \renewcommand{\arraystretch}{1.2} 
    \begin{tabular}{ccccccccccc}
        \hline
        Gmon bias& Quantities & Symbols & Units & Mode1 & Mode2 & Mode3 & Mode4 & Mode5& Mode6 & Mode7\\
        \hline 
        \multirow{1}{*}{$V_g = -0.9$}
        &Intrinsic decay rate & $\gamma_q/2\pi$ & kHz & 19.6  &   21.8   & 31.8 & 12.4  &  12.2  &  12.2   &   17.6 \\
        \hline 
        \multirow{2}{*}{$V_g = -3.0 $}
        &Coupling Strength &  $ g_m/2\pi$   &  MHz&    -0.18  &   -0.21   &  0.18 &  0.23 &   -0.18     &   -0.28     &   0.09  \\
        &Total decay rate &  $\gamma_q^{\prime}/2\pi$  &  kHz &  19.7  &  21.9  & 31.9  & 12.5  &    12.4    &     12.3   &  17.7   \\
        \hline 
        \multirow{2}{*}{$V_g = 0.75 $}
        &Coupling Strength &  $ g_m/2\pi$   &  MHz&   0.59   &  1.51  &  -0.75 & -1.52  &   0.82   & 1.67  &  -0.47 \\
        &Total decay rate &  $\gamma_q^{\prime}/2\pi$  &  kHz &   22.0 &  23.9   & 37.3  &  15.6 &    17.9    &   13.7     &  19.3   \\
        \hline
    \end{tabular}
    \label{tab2}
\end{table*}
From Eq.~\eqref{P1 of qubit}, as $g_m$ approaches zero in the weak couping regime, the exponential
term simplifies to $-\gamma_q^{\prime}t$. In the strong coupling regime, when $g_m \gg \gamma_q^{\prime}, \kappa_m $, the effects of the oscillatory term in the exponential can be neglected.
However, when $g_m$ is comparable to $\left|\gamma_q^{\prime} - \kappa_m \right|$, i.e., in the transition-coupling regime, the system exhibits distinct dynamics.

Using the insight of Eq.~\eqref{P1 of qubit} and $\gamma_q$ obtained from the coupling-off evolution, we fit the resonant evolution data for the qubit interacting with each mechanical mode under different gmon bias. The results show good agreement with our model. It is worth noting that the fitted $\kappa_m$ is larger than the value listed in Tab.~\ref{tab2}. This discrepancy arises because the phonon population in the SAWR is higher during the AC Stark measurement, leading to an increased cavity quality factor \( Q_m \)~\cite{manenti2016}. The coupling strengths are shown in Fig~\ref{fig4}(h). It shows the mode-dependent coupling strengths of the qubit at the position of $L_c/2$, following a sinusoidal variation pattern~\cite{moores2018}:
\begin{align}\label{eq_13}
    g_m = g_0\,\text{sin}\left(\frac{\pi}{2}m + \phi_q\right),
\end{align}
where $g_0$ represents an overall coupling strength. Eq.~\eqref{eq_13} exhibits sinusoidal modulation with a period corresponding to seven modes, and $\phi_q$ denotes the overall phase shift resulting from a slight deviation of the qubit position from $L_c/2$.

It is important to note that the coupling strength obtained through the fitting of the resonant evolution only provides its absolute value, $|g|$. The chosen value of the gmon bias determines the relative signs of the coupling strengths at \(V_g = -3.0\) and \(V_g = 0.75\).

\begin{figure}[!h]
    \centering
    \includegraphics[scale = 0.78]{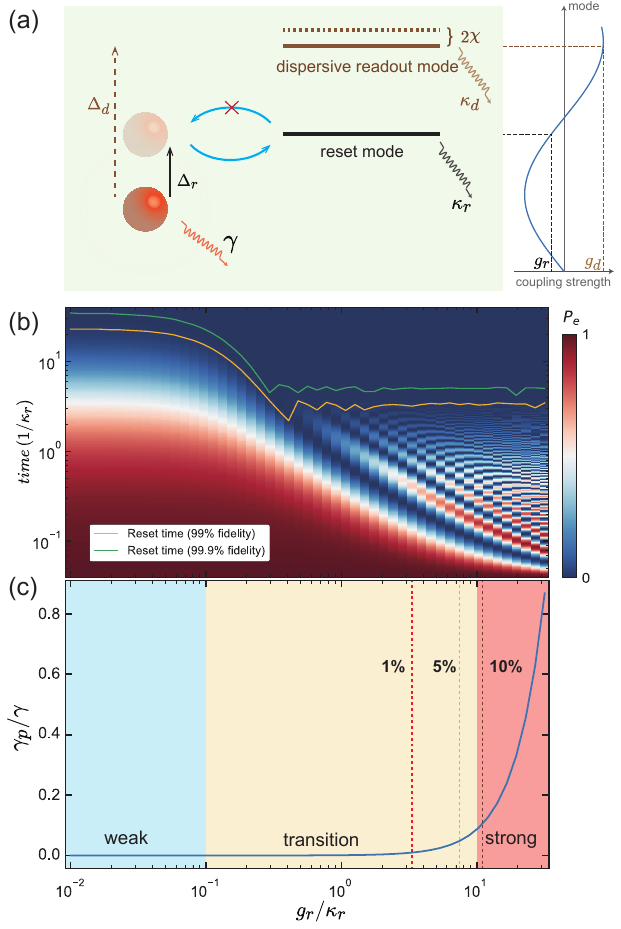}
    \caption{Intrinsic reset and readout protocol. (a) The two modes of the SAWR are utilized as the dispersive readout mode and the fast reset mode. During the reset process, the qubit is biased into resonance with the reset mode via an external magnetic flux and then decays to the ground state via the reset mode. The coupling strengths between the qubit and the reset mode, as well as the dispersive readout mode, can be engineering through their sinusoidal dependence on the mode index. (b) Numerical simulation results of the $|e\rangle$ state population of qubit during the resonant evolution with the reset mode under different coupling strengths. The yellow line in the figure represents the reset time required to achieve 99\% fidelity, while the green line corresponds to 99.9\% fidelity. As the coupling strength increases, the reset efficiency improves, reaching its minimum reset time in the transition-coupling regime. (c) The dependence of the reset mode's impact on qubit decoherence as a function of coupling strength. In the weak and transition-coupling regimes, this effect is negligible, whereas in the strong coupling regime, it becomes increasingly significant. The red, yellow, and brown dashed lines in the figure indicate the values of \( g_r / \kappa_r \) corresponding to \( \gamma_p / \gamma \) ratios of 1\%, 5\%, and 10\%, respectively.}
    \label{fig5}
\end{figure}
\section{Intrinsic Reset Protocol in Transition-Coupling Regime}

Fast reset of qubits—one of the DiVincenzo criteria~\cite{divincenzo2000}—has seen a series of advances in recent years. One major approach involves resonantly coupling the qubit level to be reset with an external dissipation source (such as the qubit’s readout cavity or a Purcell filter) via a microwave field, achieving rapid population transfer and dissipation to the ground state~\cite{magnard2018,zhou2021a,wang2024c}. However, this method requires precise calibration of multiple parameters, increasing the calibration overhead on large-scale integrated chips.

In this section, we propose a method that leverages the multimode characteristics of the mechanical cavity. One mode facilitates fast qubit reset, while another enables dispersive readout, leading to a more compact and scalable architecture for superconducting quantum chips. Through numerical simulations, we demonstrate that the coupling between the mechanical mode and the qubit achieves optimal performance in the transition-coupling regime.

Label the two modes of the SAWR as the dispersive readout mode and the reset mode, with frequencies \( \omega_d \) and \( \omega_r \), dissipation rates \( \kappa_d \) and \( \kappa_r \), and coupling strengths with the qubit \( g_d \) and \( g_r \), respectively. As shown in Fig.~\ref{fig5}(a), the reset process is implemented by applying a fast Z-bias to bring the qubit into resonance with the reset mode. Coherent evolution then proceeds until the remaining excitation reaches a predefined threshold. This duration is recorded as the reset time. According to the conclusions in Sec.~\ref{sec5}, the dynamics of the qubit and mechanical mode in the resonant regime is governed by the interplay between the coupling strength and dissipation rate. A weak dissipation rate leads to a slow relaxation to the ground state, while excessively strong coupling induces pronounced oscillations, allowing photons to cycle back into the qubit and ultimately hindering the reset efficiency.

We use QuTiP~\cite{johansson2012,johansson2013} for numerical simulations, and input experimentally relevant parameters: \( \gamma/2\pi = 0.2\,\text{MHz}\) and \( \kappa_r/2\pi = 2.5\,\text{MHz}\). The qubit is initialized in the \( |e\rangle \) state, and we vary the ratio of the coupling strength \( g_d \) to the reset mode dissipation rate \( \kappa_r \) over the range of \( 10^{-2} \) to \( 10^{2} \) to study the resonant evolution. Define the reset threshold as the population of the \( |g\rangle \) state. The reset time is defined as the time after which the population of the \( |g\rangle \) state always remains above the threshold. Fig.~\ref{fig5}(b) presents the simulation results, illustrating the reset time for different fidelity thresholds. In the weak coupling regime, the dissipation of the mechanical mode dominates the reset process, requiring a duration of at least \( 10/\kappa \) to achieve the desired threshold. As the coupling strength increases, the reset time gradually decreases, and coherent oscillations begin to emerge. After reaching a minimum, the reset time stabilizes, with the minimum occurring around \( g_m \sim \kappa_m \). These results indicate that to ensure an efficient reset rate, the coupling strength must at least reach the transition-coupling regime. 

On the other hand, we need to examine the impact of the reset mode on the qubit at the idle point, which is primarily manifested through the Purcell effect, $\gamma^{\prime} = \gamma + \gamma_p, \gamma_p = g_r^2\kappa_r/\Delta_r^2$. We calculate the dependence of \( \gamma_p / \gamma \) on \( g_r \), set $\Delta_r/2\pi = 300\,\text{MHz}$, with the results shown in Fig.~\ref{fig5}(c). In the weak and transition-coupling regimes, the Purcell effect contributes negligibly to qubit decoherence. As the system enters the strong coupling regime, the influence of the Purcell effect becomes significant, reaching approximately 10\% at the strong coupling boundary (\( g_r \sim 10\kappa_r \)). Considering both the dissipation rate and the Purcell effect, the reset mode should be coupled to the qubit in the transition-coupling regime, where the reset process is maximized in speed while the decoherence impact on the qubit remains negligible.

For the dispersive readout mode, an estimate based on the optimal condition for dispersive readout, \( 2\chi = \kappa_d \), suggests that achieving an obvious dispersive shift requires operation in the strong coupling regime with the qubit~\cite{cqed}. Due to this constraint, the intrinsic readout and reset experiment proposed in this section was not implemented in the multimode SWAR used in this work. In future designs, we can leverage the sinusoidal dependence of coupling strength on mode index to engineer the system such that the dispersive readout mode operates in the strong coupling regime, while the reset mode remains in the transition-coupling regime, as illustrated in Fig.~\ref{fig5}(a).
Another solution is to design the mechanical mode frequencies below the qubit’s idle point, utilizing the straddling dispersive coupling regime to improve qubit readout~\cite{inomata2012,boissonneault2012a}.

Using a mechanical cavity for readout not only leverages its high mode density and quality factor but also allows integration with optical fibers for optical readout. Recent works~\cite{arnold2025a,vanthiel2025} have demonstrated that optical readout can simplify cryogenic systems by eliminating components such as circulators and isolators. It can also reduce the thermal load on a single optical fiber by three orders of magnitude, offering the potential for modular scalability and parallel readout.

\section{Conclusion}
In this work, we have systematically explored the integration of a superconducting transmon qubit with a seven-mode SAWR on a chip which is fabricated with flip-chip technology, establishing a versatile platform for studying cQAD. We investigated the rich physics governing the interplay between coherent and dissipative processes in hybrid quantum systems. To this end, we combined three complementary approaches. First, we performed structural verification through phononic oscillation experiments. Second, we characterized phonon-qubit interactions spectroscopically in the dispersive regime. Third, we analyzed the dynamics of energy exchange in the transition-coupling regime. The observation of mode-dependent transitions between underdamped coherent oscillations and overdamped decay—controlled by coupler bias or mode number—reveals the important role of tunable coupling in mediating energy flow and dissipation. Our demonstration of a multimode SAWR as a dual-functional resource—enabling both dispersive readout and fast qubit reset—highlights the practical advantages of the transition-coupling regime. Future work will focus on further improving the quality factor of mechanical modes and leveraging the properties of mode-number-dependent coupling modulation to engineer a multimode SAWR with alternating strong and transition-coupling regimes.

Our findings not only underscore the practical utility of the transition-coupling regime but also offer new insights for optical readout schemes with piezo-optomechanical transducers. Such insights are important for the future design of large-scale superconducting quantum chips, laying a solid foundation for the engineering of robust and controllable quantum devices.

\textit{Note added.}——While preparing this manuscript, we became aware of an independent study~\cite{chen2025a}. The authors observe the coherent evolution between the qubit and the modes of a multimode mechanical resonator. Our work complements this study by defining the transition-coupling regime, while additionally demonstrating tunable coupling and enabling the observation of phononic responses.
\begin{acknowledgments}
This work was supported by the National Natural Science Foundation of China (Grant
Nos. 92365209, 12204528, 92265207, T2121001), the Innovation Program for Quantum Science and
Technology (Grant No. 2021ZD0301800) and the Micro/nano Fabrication Laboratory
of Synergetic Extreme Condition User Facility (SECUF). Devices were made at the
Nanofabrication Facilities at the Institute of Physics, CAS in Beijing.
\end{acknowledgments}

\appendix
\section{Measurement Setup}
\label{Measurement Setup}
\begin{figure*}[]
    \centering
    \includegraphics[width=0.8\linewidth]{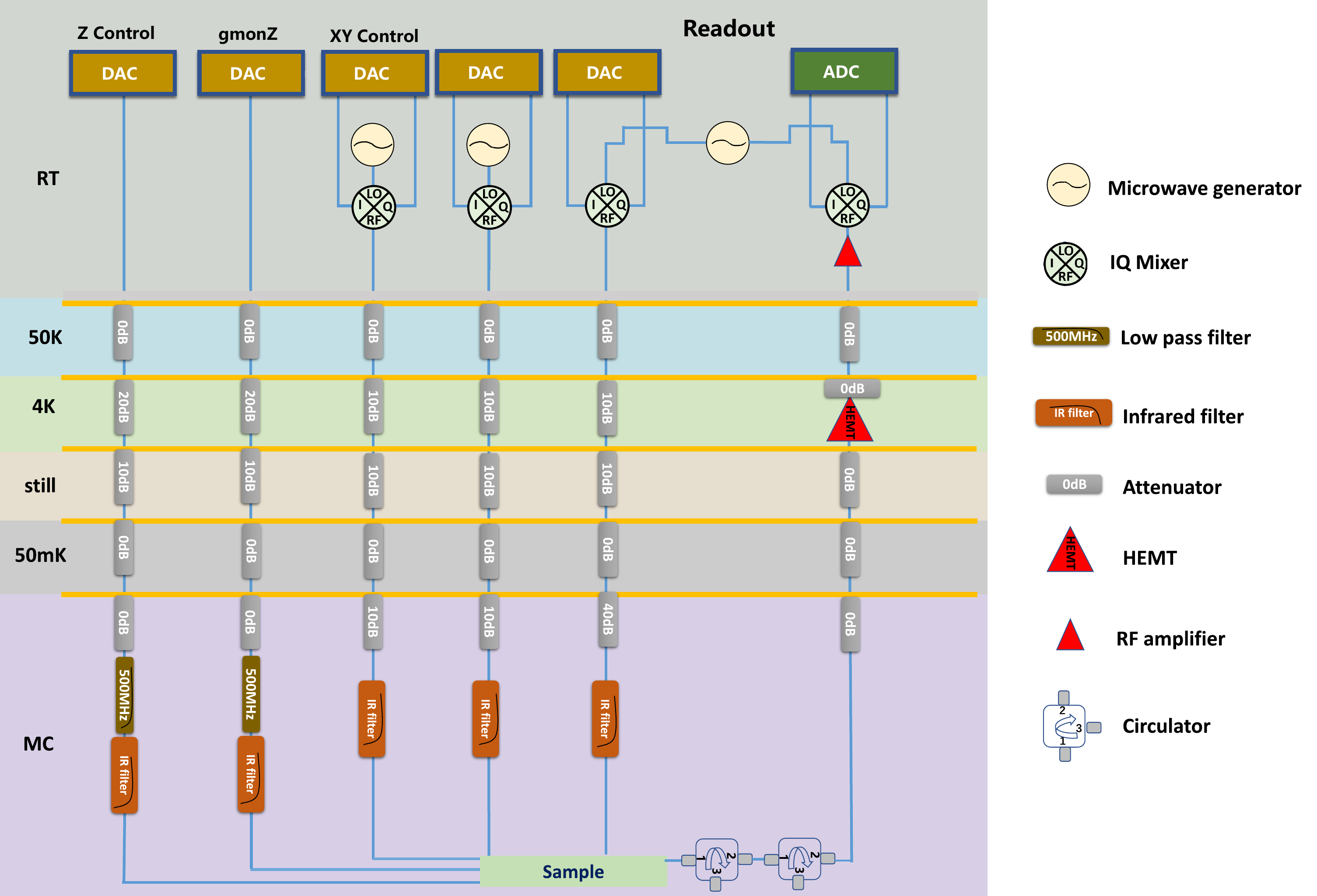}
    \caption{Measurement Setup. All measurements were performed in a dilution refrigerator (DR). The control lines were attenuated step by step to reduce thermal noise, and the readout signals were amplified in stages to improve the signal-to-noise ratio. On the right is a legend of all the components. }
    \label{fig_mesurement setup}
\end{figure*}
The experimental setup, illustrated in Fig.~\ref{fig_mesurement setup}, consists of several modules arranged sequentially from left to right: the qubit Z control, the gmon Z control, the qubit XY control, the SAW pump, and the qubit readout (including input and output). The circuitry incorporates a series of attenuators, filters, circulators, and amplifiers to ensure proper signal handling.
The qubit Z control line delivers long Z square pulses and fast pulses to tune the qubit's transition frequency. Similarly, the gmon Z control line provides long Z square pulses and fast pulses to modulate the coupling strength between the qubit and the SAW. The XY control line applies microwave pulses for manipulating the qubit state. The SAW pump line drives the interdigitated transducer (IDT) with microwave pulses, enabling the conversion of microwave photons into phonons.
For qubit readout, the qubit is capacitively coupled to a quarter-wave coplanar waveguide resonator with a frequency of $6.7005\,\text{GHz}$ (readout cavity). The opposite end of the readout resonator is inductively coupled to a transmission line. A microwave pulse at the resonator's bare frequency is injected into the transmission line and subsequently amplified by a high-electron-mobility transistor (HEMT) amplifier and a room-temperature microwave amplifier. Finally, the output signal is demodulated using an IQ mixer and digitized by an analog-to-digital converter (ADC).
\section{Phonon Oscillation}
\label{Phonon Oscillation}
\begin{figure}[t]
    \centering
    \includegraphics[width=1\linewidth]{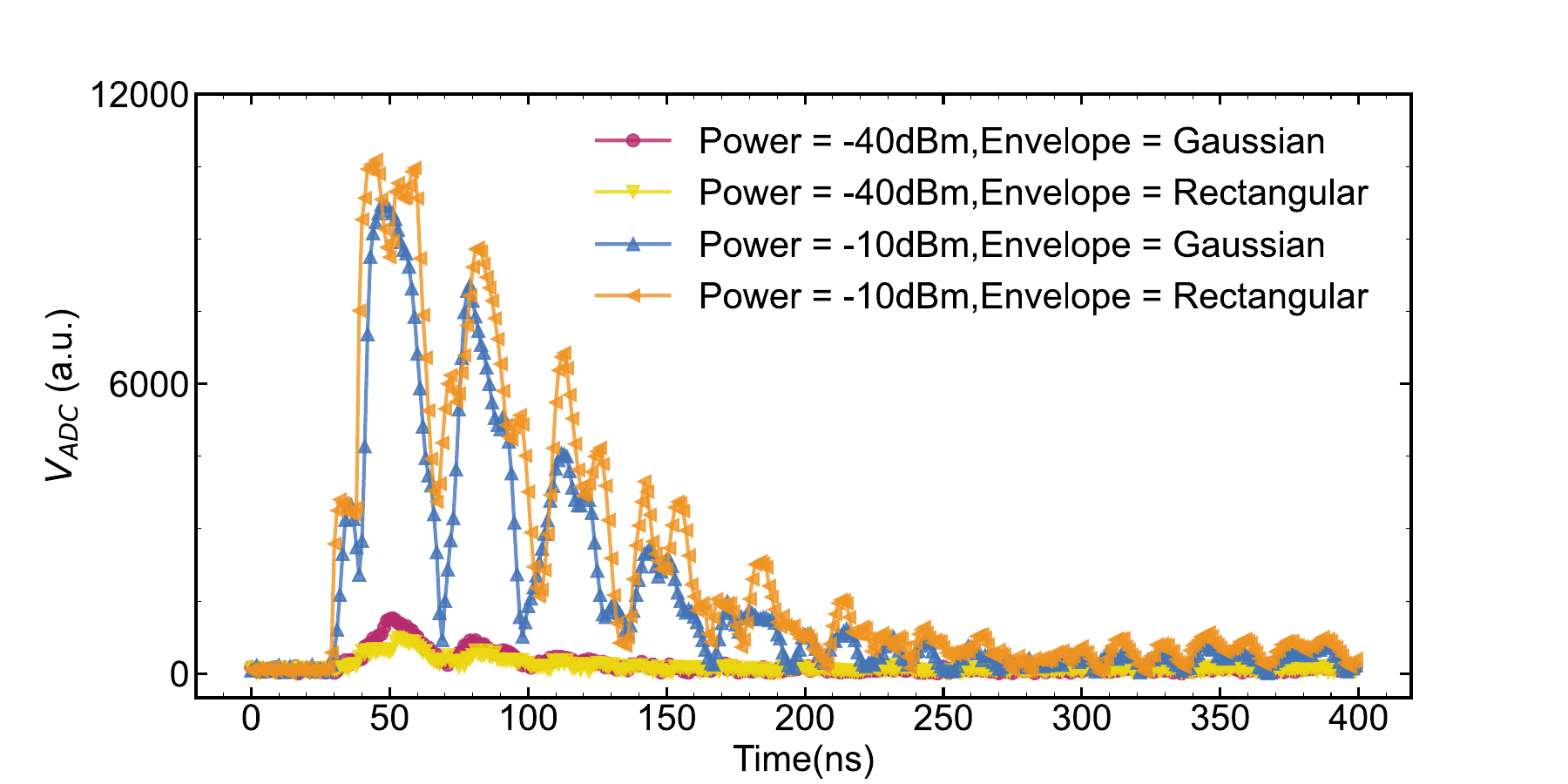}
    \caption{Phonon oscillations under different pulse powers and envelopes.}
    \label{figure_ossilation envelope}
\end{figure}
\begin{figure}[t]
    \centering
    \includegraphics[width=1\linewidth]{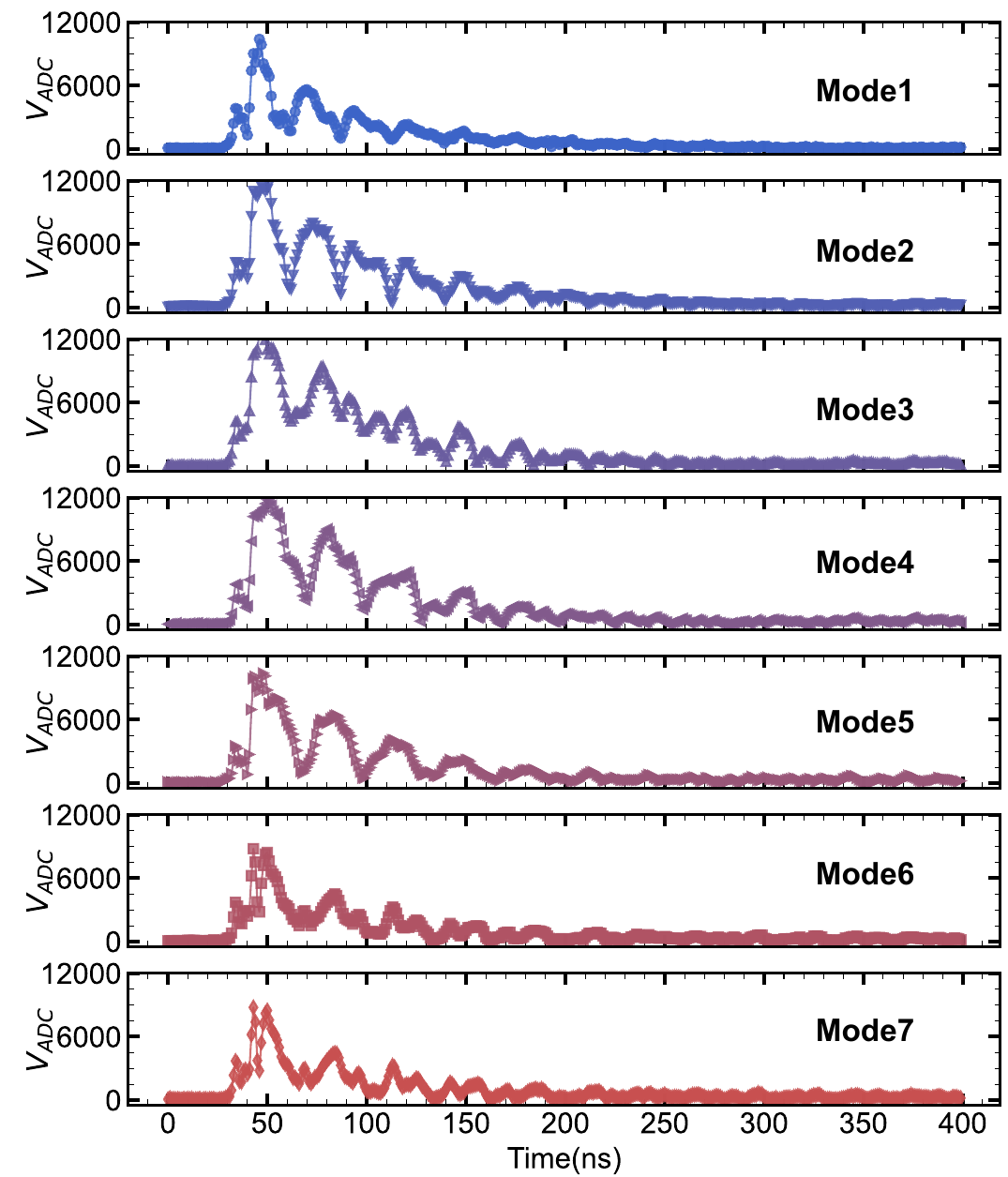}
    \caption{Experimental results of phonon oscillations of different modes.}
    \label{figure_ossilation mode}
\end{figure}

Figure~\ref{figure_ossilation envelope} presents the experimental results of phonon oscillations under different power levels and pulse envelopes, with all pulses having a duration of $12\,\text{ns}$. It can be observed that, compared to the Gaussian envelope, the oscillations induced by the rectangular envelope do not reach zero at the minima, and the oscillation peaks also differ. This discrepancy is attributed to spectral impurity in the phonon signals generated by the IDT, likely caused by pulse broadening. Therefore, Gaussian pulses were used in all experiments discussed in the main text. Figure~\ref{figure_ossilation mode} shows the phonon oscillation results for the excitation of seven modes using Gaussian pulses with a length of $12\,\text{ns}$.
\section{Frequency Calibration of Qubit}
\label{Frequency Calibration of Qubit}
\begin{figure}[t]
    \centering
    \includegraphics[width=1\linewidth]{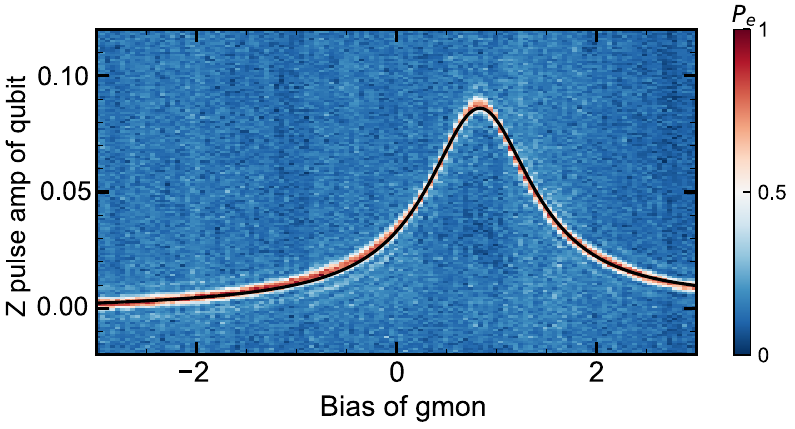}
    \caption{Experimental and fitted curves for qubit frequency calibration with different gmon bias.}
    \label{figure_freq calibration}
\end{figure}

In the experiment to determine the coupling point, it is necessary to compensate for the shift in qubit frequency at each point on-site, ensuring that the qubit frequency remains stationary at the idle point. Experimentally, this can be achieved by scanning the zpa offset and gmon bias, as depicted in Fig.~\ref{figure_freq calibration}. The relationship between the offset Z-pulse amplitude and the gmon bias can be determined by fitting the curve.

\section{Decoherence of qubit}
\label{Decoherence of qubit}
\begin{figure}[t]
    \centering
    \includegraphics[width=1.0\linewidth]{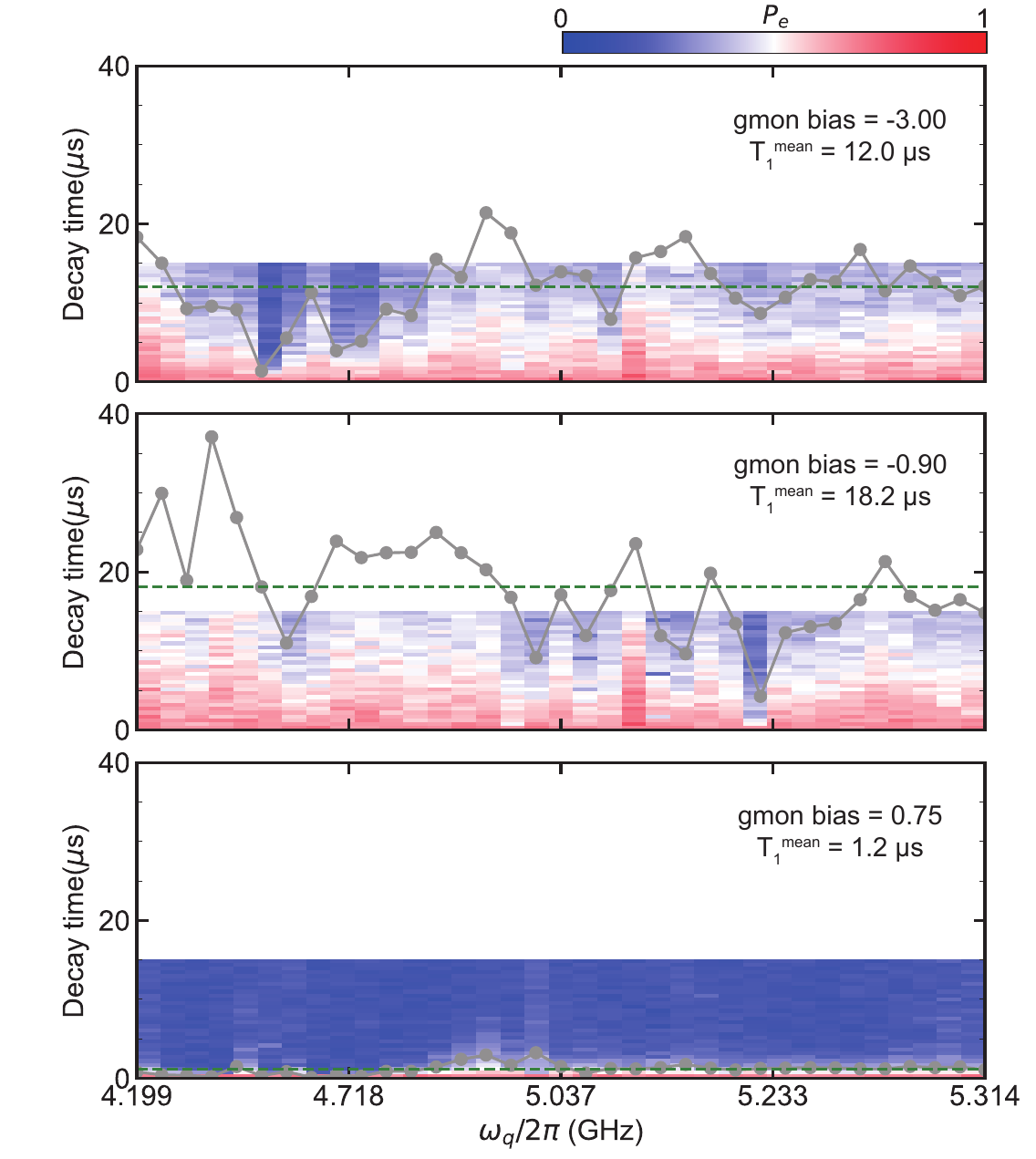}
    \caption{Two-dimensional \( T_1 \) of the qubit under different gmon bias. The gray dots represent \( T_1 \) values obtained through exponential decay fitting \(P_e = A \cdot\text{exp}(-t/T_1) + B\), while the green dashed line indicates the \( T_1^{\text{mean}} \) over the entire frequency range.}
    \label{figure_decoherence 2d}
\end{figure}

We measured the qubit lifetime $T_1$ over the frequency range from $4.2\,\text{GHz}$ to $5.3\,\text{GHz}$ for \( V_g = -3.0, V_g = 0.75, \text{and}\,V_g =-0.9\), respectively. The results are plotted in Fig.~\ref{figure_decoherence 2d}. The average lifetime of the qubit reaches a maximum of approximately $18.2\,\text{$\mu s$}$ when the coupling is turned off. At maximum coupling, the average lifetime decreases to $1.2\,\text{$\mu s$}$.
\begin{figure}[t]
    \centering
    \includegraphics[width=0.95\linewidth]{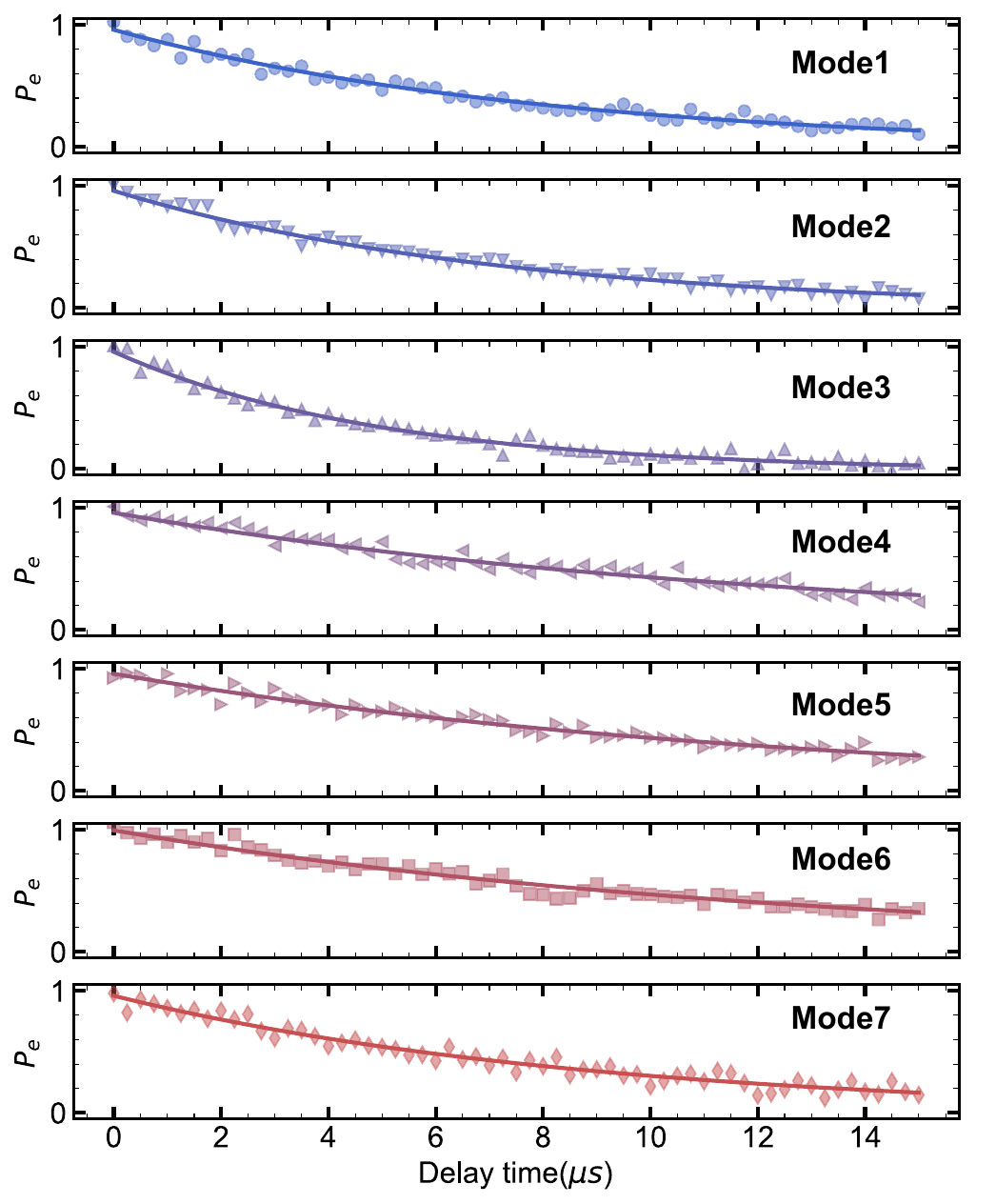}
    \caption{The $P_e$ of the qubit during the time evolution at the frequency of mechanical modes, when $V_g = -0.9$.}
    \label{figure_resonate coupling off}
\end{figure}

To determine the intrinsic dissipation rate of the qubit when it is resonant with the seven mechanical modes, we measured the lifetime of the qubit at the corresponding resonance frequencies with the coupling turned off. The data are presented in Fig.~\ref{figure_resonate coupling off}. It can be observed that the intrinsic dissipation rates of the qubit at these seven positions are very similar, indicating the absence of non-Markovian dissipation at these points. This ensures that our resonant evolution model can successfully describe the qubit and multimode SAW resonator composite system.

\clearpage
\bibliography{cQAD_in_transition_coupling_regime}

\end{document}